# VIVA PANSPERMIA!

Chandra Wickramasinghe
Cardiff Centre for Astrobiology
Cardiff University, Cardiff UK

The arguments for panspermia as a mode of origin of life on Earth are far from dead; on the contrary they are now more robust than ever.

The term *panspermia* is derived from Greek roots: *pan* (all) and *sperma* (seed) –seeds everywhere. The underlying ideas go back to the time of classical Greece to philosopher Aristarchus of Samos in the 5$^{th}$ century BC promulgating the ominipresence of the seeds of life in the cosmos. Panspermia also has a resonance with more ancient Vedic, Hindu and Buddhist traditions of India stretching back over 4000 years.

The first serious scientific statement of panspermia came from Lord Kelvin (William Thomson[1]) at the 1881 presidential address to the British Association:

*"…Hence, and because we all confidently believe that there are at present, and have been from time immemorial, many worlds of life besides our own, we must regard it as probable in the highest degree that there are countless seed-bearing meteoritic stones moving about through space. If at the present instant no life existed upon the Earth, one such stone falling upon it might, by what we blindly call natural causes, lead to its becoming covered with vegetation."*

Two decades later panspermia was placed in an explicit astronomical context by Svante Arrhenius[2], first in a short paper published in 1903 and thereafter in his book *Worlds in the Making*. Long before the discovery of the many remarkable survival attributes of bacteria, Arrhenius inferred that such properties must exist, and cited experiments where seeds had been taken down to near zero degrees Kelvin and shown to survive. Arrhenius also calculated the effect of radiation pressure of starlight on spore-sized particles in space and argued that spores lofted in rare events from an inhabited planet like Earth could be projected at speed to reach a distant planetary system. This came at a time when neo-Darwinian ideas of evolution were at last beginning to gain general support, and it might have been feared that Arrhenius' views would threaten or reverse a hard-won victory over creationist beliefs. The threat to Darwinism was unfounded however. Darwin did not make any reference whatever to the origin of life in his classic book *Origins of Species*, although in a letter to Joseph Hooker in 1871 he wrote thus:

*"But if (and oh! what a big if!) we could conceive in some warm little pond, with all sorts of ammonia and phosphoric salts, light, heat, electricity, &c., present, that a proteine compound was chemically formed and ready to undergo still more complex changes….."*

That conjecture did not necessarily relate to a terrestrial origin of life, although it was later interpreted as such, and formed the inspiration for the familiar primordial soup model and to theories of chemical evolution that became more or less adjunct to the Darwinian theory.

Opposition to Arrhenius's challenge of Earth-bound theories of life's origin took a ferocious turn with publications by Becquerel[3] and others claiming to *disprove* panspermia from an experimental standpoint. On the basis of experiments that showed certain bacteria to be killed by exposure to ultraviolet light, the argument gained ground that *all* bacteria expelled from a planet would be killed by conditions in space. It is now clear that space-travelling bacteria could be easily shielded from ultraviolet light with extremely thin layers of overlying carbon, and bacteria within interiors of small clumps would be particularly well protected[4-6]. This was not known at the time, and in any case, as Julius Caesar said:

*fere libente homines id quod volunt credunt – men readily believe that they want to believe*

Thus a firm conviction that panspermia is a defunct theory gained ground and dominated scientific culture for nearly half a century from 1924 – 1974.

The revival of panspermia as a viable theory started in the mid-1970's with the work of Hoyle and the present author[7-9] seeking to explain the steadily increasing complexity of the organic molecules that were being discovered in interstellar space with the use of radio astronomy and infrared astronomy. At first, rather subtle data reduction methods were required to infer the presence of complex organics from the earliest infrared spectra of dust[9]. Both Hoyle and the author devoted nearly 5 years of our professional lives to this project, and by 1983 we inferred confidently that some 30 percent of the carbon in interstellar dust clouds had to be tied up in the form of organic dust that matched the properties of degraded or desiccated bacteria[4]. This far-reaching conclusion has only come to be further strengthened with advances in stellar spectroscopy in the past three decades. Spectroscopic signatures of PAH's and organic polymers in interstellar space as well as in external galaxies have come to be well established[10-14].

What cannot be denied is that the Hoyle-Wickramasinghe corpus of published work comprised of several books and hundreds scientific papers seeking to re-establish panspermia as a viable theory on a proper scientific footing came well ahead of the references that are often cited[15,16]. Recent studies of extremophiles and the well-established resistance of bacteria to ionising radiation[17] and space conditions including hypervelocity impacts[18] lend further credibility to panspermia theories. In a recent review of relevant experimental data on this subject we concluded that despite all the hazards of space a minute fraction of bacteria *must* remain in a viable condition in interstellar clouds between expulsion from one planetary/cometary source and re-accommodation in another[6, 19]. For panspermia to work this viable fraction could be as small as 1 in $10^{24}$ a condition that would be well nigh impossible to violate. The picture here is strikingly similar to the sowing of seeds in the wind. Few are destined to survive, but so many are the seeds that some amongst them would inevitably manage to take root.

Hoyle and the present author in our writings have elaborated on the role of partially destroyed bacteria noting that viral genomes derived from cells have a much longer persistence under interstellar conditions compared to the much larger bacterial (or eukaryotic) genome. In our monograph[4] *Proofs the Life is Cosmic* published in 1982 we wrote thus (p.14):

*Viruses, and viroids still more, have the advantage of being smaller targets for damaging radiation than bacteria. Thus about 100kr (of ionising radiation) is needed to produce a single break in the nucleic acid of the smaller viruses, and in excess of 1Mr for viroids. In addition to this advantage, viruses can use the enzymic apparatus of host cells to repair themselves, even to the astonishing extent of being able to "cannibalise", a process in which several inactivated viral particles combine portions of themselves to produce a single active particle.*

The currently popular view that all the organics now known to exist in interstellar clouds represent steps towards life is not justified. We have recently reviewed the astronomical evidence that supports the idea these molecules are most probably derived from life[14,20] - the interstellar medium is a veritable graveyard of cosmic bacteria. The detritus of bacterial life in interstellar clouds would range from charred bacteria (resembling anthracite grains), genetic fragments of cells representing viruses and viroids, to PAH's and smaller organic molecules.

The present author's views do not contradict Wesson's restatement of this process as *necropanspermia* – cute term indeed! However, it is impossible to maintain, as he does, that *all bacteria* expelled from a source will be killed in interstellar transits. Indeed explicit mechanisms for viable interstellar transfers of microorganisms have been identified by several authors[21,22]. The first introduction of life onto our planet (or indeed any planet) must involve the introduction of a viable microorganism – not a fragment of genome, a virus or a dead microorganism. Subsequent evolution of life would be greatly speeded up with the more prolific injection of viruses that could insert genes into already evolving cells. We have argued that this process could not only lead to epidemics of disease but also contribute to evolution[13,23,24].

An initial injection of a viable cellular life form, which takes root and begins to evolve, would be augmented genetically by viruses carrying genes for the development of all other possible life forms [4,23,24]. This grand ensemble of genes for cosmic evolution would in our model have been delivered in comet dust to our planet throughout geological time[4]. The earliest evidence of bacterial life on the Earth is between 3.8 and 4 bya during the Hadean epoch[25] which was characterised by an exceptionally high rate of comet impacts. Comets that delivered water to form most of the oceans probably delivered the first viable bacterial cells that subsequently evolved. From an initial small bacterium (typified by *Mycoplasma genitalium*) that had ~500 genes, life evolves over a 4 billion year timescale to produce mammals with genomes consisting of some 25,000 genes. Modelling the correlation between average gene number and time elapsed leads to an empirical relation

$$N \cong 500 \exp(t/\tau) \qquad (1)$$

where the value of $\tau$ is close to 1by (see Sharov[26]; Joseph and Wickramasinghe, in press). Equation (1) may be taken as defining the development of gene complexity within a physically connected set of planets, $dN/dt \propto N$ implying a capture rate of genes proportional to cross-section in an open Darwinian system of evolution.

Working backwards from $N=500$ at 4 bya to lead to a simple viral-sized genome of say $N = 10$ (the virus of *E coli* φX174 has 11 genes), nearly four e-folding times are involved, giving a total evolutionary timescale of nearly 8 billion years, longer than the age of the Earth (cf, Joseph[28], Joseph and Schild[29]).

Hoyle and the present author have dwelt at length on the improbability of obtaining a minimal gene set needed for the emergence of a bacterial genome from random processes. For a set of 500 genes and assuming that 10 sites per gene need to be correctly filled with one of a set of 20 amino acids, the probability turns out to be $\sim 10^{-6500}$. But if only a set of 60 genes can kick start the evolutionary process in the early universe, leading eventually the genes to all life, then the probability is $10^{-80}$, which is more easily attainable, and could have been achieved in situations such as have been discussed by Gibson et al[27].

*Viva panspermia!*